\documentclass[draftclsnofoot, onecolumn]{IEEEtran}
\IEEEoverridecommandlockouts
\usepackage{cite}
\usepackage{amsmath,amssymb,amsfonts}
\usepackage{algorithmic}
\usepackage{graphicx}
\usepackage{textcomp}
\usepackage{xcolor}
\usepackage{tabularx}
\usepackage{datatool}
\usepackage{float}
\usepackage{booktabs}

\def\BibTeX{{\rm B\kern-.05em{\sc i\kern-.025em b}\kern-.08em
    T\kern-.1667em\lower.7ex\hbox{E}\kern-.125emX}}

\begin{document}


\thispagestyle{empty}

\newcolumntype{L}[1]{>{\raggedright\arraybackslash}p{#1}}
\newcolumntype{C}[1]{>{\centering\arraybackslash}p{#1}}
\newcolumntype{R}[1]{>{\raggedleft\arraybackslash}p{#1}}

\clearpage
\pagenumbering{arabic} 

\title{Cyber Value At Risk Model for IoT Ecosystems}

\makeatletter
\newcommand{\linebreakand}{
  \end{@IEEEauthorhalign}
  \hfill\mbox{}\par
  \mbox{}\hfill\begin{@IEEEauthorhalign}
}

\makeatother
\author{
    Goksel Kucukkaya\textsuperscript{1}, 
    Murat Ozer\textsuperscript{1}, 
    Emrah Ugurlu\textsuperscript{2}\\[1em]
    \textsuperscript{1}\textit{School of Information Technology, University of Cincinnati, Cincinnati, Ohio, USA} \\
    \textsuperscript{2}\textit{Ostim University, Ankara, Turkiye} \\[1em]
    kucukkgl@ucmail.uc.edu, m.ozer@uc.edu, emrah.ugurlu@ostimteknik.edu.tr 
}

\maketitle

\thispagestyle{plain}
\pagestyle{plain}

\begin{abstract}
The Internet of Things (IoT) presents unique cybersecurity challenges due to its interconnected nature and diverse application domains. This paper explores the application of Cyber Value-at-Risk (Cy-VaR) models to assess and mitigate cybersecurity risks in IoT environments. Cy-VaR, rooted in Value at Risk principles, provides a framework to quantify the potential financial impacts of cybersecurity incidents. Initially developed to evaluate overall risk exposure across scenarios, our approach extends Cy-VaR to consider specific IoT layers: perception, network, and application. Each layer encompasses distinct functionalities and vulnerabilities, from sensor data acquisition (perception layer) to secure data transmission (network layer) and application-specific services (application layer). By calculating Cy-VaR for each layer and scenario, organizations can prioritize security investments effectively. This paper discusses methodologies and models, including scenario-based Cy-VaR and layer-specific risk assessments, emphasizing their application in enhancing IoT cybersecurity resilience. 
\end{abstract}
 
\section{Introduction}
One of the latest definitions of the Internet of Things (IoT) states that IoT devices:

1. Possess at least one transducer (sensor or actuator) for interacting directly with the physical world.
2. Include at least one network interface.
3. Are not conventional Information Technology devices, such as smartphones and laptops, for which the identification and implementation of cybersecurity features are already well understood.
4. Can function independently and are not solely components of other devices, such as processors \cite{IoTCyberSecurityImprovementAct2020}.

The security of Internet of Things (IoT) devices poses significant challenges due to their widespread deployment, diverse functionalities impatcing business outcomes, and varying levels of resource constraints. These challenges are further exacerbated by the often-limited computational and memory capabilities of IoT devices, which make the implementation of robust security measures difficult \cite{muhammad2015critical}. Ensuring the security of IoT systems is crucial to protect against unauthorized access, data breaches, and other cyber threats that can exploit vulnerabilities in IoT networks \cite{Neshenko2019_IoTSecuritySurvey}.

In response to these security challenges, IoT architecture has emerged as a critical framework for both developing secure IoT solutions and analyzing existing systems. An effective IoT architecture provides a structured approach to integrating security features across all layers of IoT systems, from devices and gateways to cloud services and applications. This architectural framework facilitates the implementation of standardized security protocols, encryption mechanisms, and authentication methods, thereby enhancing the overall security posture of IoT deployments \cite{Gubbi2013_IoTArchitecture}. These architectures also help simplify and abstract reality into human comprehendible models.

\section{IoT Architectures}
In this section, we visited IoT architectures that evolved to address the changing landscape of security threats and technological advancements used in and outside the IoT ecosystem. 
\subsection {3-layer architecture}
The three-layer architecture is one of the fundamental and widely adopted IoT architectures. It is known for its simplicity and ease of implementation, making it accessible for various IoT applications. The architecture comprises three main layers: the perception layer, the network layer, and the application layer \cite{Burhan2018-dl}.

The perception layer, situated at the base of the traditional three-layer IoT architecture, serves as the foundation for hosting smart devices.

The network layer functions as the central hub for data forwarding tailored to specific services. It facilitates secure communication among smart devices and with the cloud using IoT gateway technologies like wired, Wi-Fi, and cellular networks. This layer also ensures each system possesses unique addressing and routing capabilities, enabling seamless integration of an extensive array of devices into a unified network.

The application layer is responsible for delivering customized, application-specific services to users. These services span diverse domains such as smart education, healthcare, energy grids, transportation, and more.

\subsection{4-layer architecture}
A four-layer architecture was introduced as an advancement over the basic three-layer architecture, which could not fully meet all the evolving requirements of IoT due to ongoing developments in the field. This newer architecture retains the three foundational layers of its predecessor—perception, network, and application layers—while adding a fourth layer known as the support layer.  Management Capabilities, such as QoS Manager, Exchange and Management, Trust and Reputation, Identity Management and Authentication and Device Manager are part of the support layer. \cite{darwish2015improved}. 

\subsection{ 5-layer architecture}
The five-layered architecture adds a business layer that oversees the management of the entire IoT system, including applications and services. It develops business models, graphs, flowcharts, and other structures based on data received from the Application layer. The effectiveness of IoT technology heavily relies on robust business models. Through analysis of outcomes, this layer plays a crucial role in shaping future actions and business strategies \cite{https://doi.org/10.1155/2017/9324035}.

As can be inferred, early IoT architectures were primarily focused on connectivity and data exchange, with limited emphasis on security. However, as the proliferation of IoT devices increased and cyber threats became more sophisticated, the need for a more robust and comprehensive architectural approach became evident. Modern IoT architectures now incorporate advanced security measures such as end-to-end encryption, secure boot processes, intrusion detection systems, and anomaly detection algorithms. These enhancements are designed to protect IoT devices and networks from a wide range of security threats, including malware attacks, data tampering, and unauthorized access. Furthermore, the evolution of IoT architecture continues to be driven by ongoing research and development efforts aimed at addressing emerging security challenges and improving the resilience of IoT ecosystems.

\section{Regulations for IoT }
Enacted in 2020, the IoT Cybersecurity Improvement Act establishes minimum security standards for IoT devices owned and controlled by the federal government. This legislation empowers the Chief Information Officer (CIO) to prohibit agency heads from procuring, renewing contracts for, or using IoT devices if a mandatory review finds that their use would prevent compliance with NIST standards and guidelines.

The CIO can grant a waiver to this requirement only under specific circumstances:
\textit{If the waiver is deemed necessary in the interest of national security,
If the procurement, use, or obtaining of such devices is essential for research purposes,
If alternative and effective security methods appropriate to the device's function are implemented.} \cite{uscongress2020}

Proponents of government regulations argue that these regulations implement uniform standards across all devices to enhance consumer security and can mandate regular patches that address emerging security threats. Opponents, however, hold a contrasting perspective. They are concerned about the potential stifling of innovation and increased bureaucratic hurdles due to regulation and costly regulatory compliance that could disproportionately affect smaller companies, limiting consumer choice and market competition.

These regulations are not mandatory, leaving CIOs with the challenge of making critical trade-offs. They must decide to what extent they should invest in IoT security measures. Balancing the need for robust security against budget constraints and operational efficiency is essential. CIOs must evaluate the potential risks and benefits of various security investments, determining the appropriate level of protection required for their specific IoT deployments. This involves considering factors such as the sensitivity of the data being handled, the potential impact of security breaches, and the cost-effectiveness of different security solutions.

\section{Decision Making in IoT Security Investment}

This issue has been addressed by researchers who have developed frameworks and guidelines to assist CIOs in making informed decisions about IoT security investments. These frameworks often include risk assessment models, cost-benefit analyses, and best practices for implementing security measures tailored to the specific needs of IoT environments. By leveraging these research-based tools, CIOs can more effectively balance the trade-offs between security, cost, and functionality, ensuring that their IoT systems are both secure and operationally efficient.

One solution has been the application of investment theory: In cybersecurity literature, significant efforts have been made to adapt the principles and metrics of investment theory to security investments. The key metric in investment theory is the ratio of cost to benefit, or, in terms of a production function, the amount of output per unit of input. To provide a structured approach, the high-level security production function is decomposed into two steps: first, the cost of security is mapped to a security level; second, this security level is mapped to the resulting benefits. This approach helps organizations evaluate and optimize their security investments by quantitatively analyzing the trade-offs between costs and expected outcomes.\cite{bohme2008economic}

Another security investment model, developed by \cite{gordon2002economics}, defines a security breach probability function. This model maps the monetary value of security investments to the probability of incurring a defined loss. By quantifying the relationship between investment and risk reduction, this model allows organizations to predict the likelihood of security breaches based on their investment levels. This approach provides a more precise and data-driven method for determining optimal security expenditures, helping organizations to allocate resources more effectively to minimize potential losses.

Return on Security Investment (ROSI) models have been extensively explored in cybersecurity research. ROSI models aim to quantify the financial return from investments in security measures. By comparing the cost of implementing security controls with the expected reduction in losses due to security incidents, these models provide a clear economic perspective on the value of security investments \cite{returrn_on_sec_investment}. ROSI helps organizations justify their security expenditures by demonstrating how effective security measures can lead to significant cost savings and risk mitigation over time.

\subsection{Cyber Value at Risk}

Based on the principles of Value at Risk (VaR) \cite{albina}, and to assist organizations facing cybersecurity challenges, the World Economic Forum’s Partnering for Cyber Resilience initiative (WEF 2012) introduced a model to measure and quantify the impact of cyber threats on businesses and their exposure to these threats. This model, known as Cyber Value-at-Risk (Cy-VaR), provides a comprehensive framework for assessing the potential financial impact of cybersecurity incidents. By leveraging VaR methodologies, Cy-VaR helps organizations understand their risk exposure and make informed decisions about their cybersecurity investments and strategies \cite{Deloitte2015}

The Cy-VaR can be represented mathematically as \cite{mcneil2015quantitative}:

\begin{equation}
\text{VaR}_{\alpha}(L) = \inf \{ l \in \mathbb{R} | P(L > l) \leq 1 - \alpha \}
\end{equation}

Where:
\begin{itemize}
    \item $\text{VaR}_{\alpha}(L)$ represents the Value at Risk at the confidence level $\alpha$.
    \item $L$ is the potential loss.
    \item $\inf$ denotes the infimum.
    \item $P(L > l)$ is the probability that the loss $L$ exceeds the threshold $l$.
    \item $1 - \alpha$ is the acceptable risk level.
\end{itemize}

It is more convenient to work with scenarios and vignettes, making table top exercises and capture the flag exercises useful in value at risk calculations. We call them here in generic forms, as scenario based value extraction.
For each scenario $i$, calculate the CyVaR as the product of its probability $P_i$ and its associated financial loss $L_i$:

\begin{equation}
\text{CyVaR}_i = P_i \times L_i
\end{equation}

Where:
\begin{itemize}
    \item $\text{CyVaR}_i$: Cyber Value at Risk for scenario $i$.
    \item $P_i$: Probability of scenario $i$ occurring.
    \item $L_i$: Financial loss associated with scenario $i$.
\end{itemize}

To obtain the total probabilistic CyVaR, sum up the individual CyVaR values across all scenarios \cite{albina}:

\begin{equation}
\text{Total CyVaR} = \sum_{i} \text{CyVaR}_i = \sum_{i} (P_i \times L_i)
\end{equation}

Incorporating IoT layers into this formula can help to focus on each layer as a separate problem area and be more specific. The key IoT layers considered are the perception layer, the network layer, and the application layer. For each IoT layer $j$ (where $j$ represents perception, network, and application) and scenario $i$, calculate the layer-specific CyVaR as:

\begin{equation}
\text{CyVaR}_{ij} = P_{ij} \times L_{ij}
\end{equation}

Where:
\begin{itemize}
    \item $\text{CyVaR}_{ij}$: Cyber Value at Risk for layer $j$ in scenario $i$.
    \item $P_{ij}$: Probability of scenario $i$ occurring in layer $j$.
    \item $L_{ij}$: Financial loss associated with scenario $i$ in layer $j$.
\end{itemize}

To obtain the total probabilistic CyVaR across all IoT layers and scenarios, sum up the individual CyVaR values:

\begin{equation}
\text{Total CyVaR} = \sum_{j} \sum_{i} \text{CyVaR}_{ij} = \sum_{j} \sum_{i} (P_{ij} \times L_{ij})
\end{equation}

This approach allows organizations to assess their overall risk exposure by considering various scenarios across different IoT layers, each with its own probability and potential financial impact.

Incorporating IoT layers into the CyVaR model provides a granular view of cybersecurity risks associated with each layer of the IoT architecture. The perception layer is responsible for hosting smart things and gathering data through sensors and actuators. The network layer enables secure communication between devices and the cloud, using technologies such as wired, Wi-Fi, and cellular networks. The application layer provides customized services to users, such as smart education, smart health, smart energy grids, and smart transportation. By evaluating the CyVaR for each layer, organizations can prioritize security investments and mitigation strategies more effectively, addressing specific vulnerabilities unique to each layer of the IoT architecture.

\section{Conclusion}

In conclusion, this paper has explored the application of Cyber Value-at-Risk (Cy-VaR) models in assessing and managing cybersecurity risks within Internet of Things (IoT) environments. By leveraging principles rooted in Value at Risk (VaR), we have demonstrated how Cy-VaR can provide a quantitative framework to estimate potential financial impacts of cybersecurity incidents across various IoT scenarios.

Our approach extends beyond traditional risk assessment by incorporating specific IoT layers—perception, network, and application—each presenting unique security challenges and vulnerabilities. By calculating Cy-VaR for these layers, organizations can prioritize security investments and tailor mitigation strategies to address specific risks effectively.

\section{Discussion and Future Directions}

In addition to the theoretical framework and practical applications discussed, future research can explore enhanced methodologies to strengthen the application of Cyber Value-at-Risk (Cy-VaR) models in IoT cybersecurity. One promising avenue is the integration of Monte Carlo simulations to assess cybersecurity risks under various probabilistic distributions. Monte Carlo simulations can simulate a wide range of scenarios by randomly sampling inputs based on probability distributions, providing a robust method to quantify uncertainty and assess the impact of cyber threats on IoT systems. This approach not only enhances the accuracy of risk assessments but also enables organizations to identify and prioritize critical vulnerabilities more effectively.

Furthermore, leveraging agent-based modeling (ABM) presents another promising direction to enhance the understanding of cybersecurity dynamics within IoT connected networks. ABM can simulate interactions between autonomous agents (representing IoT devices, users, and malicious actors) within a networked environment, allowing researchers to analyze emergent behaviors and vulnerabilities. By incorporating ABM into Cy-VaR models, researchers can simulate realistic cyber-attack scenarios, evaluate the effectiveness of security measures, and optimize response strategies in dynamic IoT ecosystems.

These advancements in simulation and modeling techniques are crucial for advancing cybersecurity resilience in IoT. By combining theoretical insights with practical simulations, future research can develop more adaptive and proactive cybersecurity strategies, mitigating risks before they manifest and ensuring the integrity and reliability of IoT deployments across diverse application domains.

\bibliographystyle{ieeetr}
\bibliography{references}

\end{document}